\documentclass[twocolumn,prl,superscriptaddress,preprintnumbers]{revtex4}[11pt]
\pdfoutput=1
\usepackage{amsmath,amssymb,graphicx} 
\usepackage{epsf,verbatim}
\usepackage{hyperref}
\usepackage{comment}
\usepackage{color}
\usepackage[normalem]{ulem}

\newcommand{\mysection}[1]{{\vskip 2mm \noindent \bf #1.--}}

\newcommand{\simgt}{\mathrel{\lower2.5pt\vbox{\lineskip=0pt\baselineskip=0pt
           \hbox{$>$}\hbox{$\sim$}}}}
\newcommand{\simlt}{\mathrel{\lower2.5pt\vbox{\lineskip=0pt\baselineskip=0pt
           \hbox{$<$}\hbox{$\sim$}}}}

\newcommand{\squishlist}{
 \begin{list}{$\bullet$}
  { \setlength{\itemsep}{0pt}
     \setlength{\parsep}{3pt}
     \setlength{\topsep}{3pt}
     \setlength{\partopsep}{0pt}
     \setlength{\leftmargin}{1.5em}
     \setlength{\labelwidth}{1em}
     \setlength{\labelsep}{0.5em} } }

\newcommand{\squishlisttwo}{
 \begin{list}{$\bullet$}
  { \setlength{\itemsep}{0pt}
     \setlength{\parsep}{0pt}
    \setlength{\topsep}{0pt}
    \setlength{\partopsep}{0pt}
    \setlength{\leftmargin}{2em}
    \setlength{\labelwidth}{1.5em}
    \setlength{\labelsep}{0.5em} } }

\newcommand{\squishend}{
  \end{list}  }

\newcommand{\be}{\begin{equation}}
\newcommand{\ee}{\end{equation}}
\newcommand{\bea}{\begin{eqnarray}}
\newcommand{\eea}{\end{eqnarray}}

\begin{document}

\title{Importance of upgraded energy reconstruction for direct dark matter searches with liquid xenon detectors} 

\preprint{}

\author{Peter Sorensen}
\thanks{pfs@llnl.gov}
\affiliation{Lawrence Livermore National Laboratory, 7000 East Ave., Livermore, CA 94550, USA} 
 
\begin{abstract} 
The usual nuclear recoil energy reconstruction employed by liquid xenon dark matter search experiments relies only on the primary scintillation photon signal. Energy reconstruction based on both the photon and electron signals yields a more accurate representation of search results. For a dark matter particle mass $m_{\chi}\sim10$~GeV, a nuclear recoil from a scattering event is more likely to be observed in the lower left corner of the typical search box, rather than near the nuclear recoil calibration centroid. In this region of the search box, the actual nuclear recoil energies are smaller than the usual energy scale suggests, by about a factor $\times2$. Recent search results from the XENON100 experiment are discussed in light of these considerations. 
\end{abstract}

\maketitle

 \setcounter{equation}{0} \setcounter{footnote}{0}

\mysection{{\it Introduction}}
\label{sec:intro}
Liquid xenon detectors are presently at the forefront of direct searches for galactic particle dark matter. They have placed the most stringent upper limits on the dark matter $-$ nucleon cross section $\sigma_n$ \cite{:2012nq}, and an additional factor $\times5$ (or more) improvement in sensitivity is expected in the coming year \cite{Akerib:2011ix}. The expected signature of a dark matter interaction with target nuclei is, in a large number of theoretical models, a low energy nuclear recoil \cite{Gaitskell:2004gd}. Liquid xenon offers an approximate factor $\times200$ discrimination between nuclear recoils (as from neutrons, and expected for dark matter), and electron recoils (as from electromagnetic background) \cite{Chepel:2012sj}. The basis for the discrimination arises from the fact that the partitioning of electronic energy losses into a number of photons ($n_{\gamma}$) and electrons ($n_e$) is a function of the incident particle type, and also of the particle energy \cite{Sorensen:2011bd}. Thus search results from liquid xenon detectors are generally given in terms of a nuclear recoil energy (${E}_{nr}$) estimator versus the discriminant, $y\propto n_e/n_{\gamma}$. 

In practice, this has exclusively taken the forms ${E}_{nr}~\propto~\mbox{S1}$ and $y~\propto~\mbox{S2/S1}$. In these equations, $\mbox{S1}~=~ \alpha_1~n_{\gamma}$ and $\mbox{S2}~=~ \alpha_2~n_e$ are the experimentally measured number of photoelectrons (from photomultiplier tubes), corresponding to the photon and electron signal.   Typical values are $\alpha_1 \approx \frac{1}{15}$ and $\alpha_2 \approx 20$, and these quantities are detector-dependent.  The purpose of this Letter is to show that an energy reconstruction based on both $n_e$ and $n_{\gamma}$ is significantly more accurate and useful than the usual ${E}_{nr}~\propto~\mbox{S1}$ method, especially in considering possible detection scenarios.

\mysection{{\it Discussion}}
\label{sec:discuss}
Nuclear recoil energy reconstruction in liquid xenon has historically been defined by ${E}_{nr}~=~\langle \mbox{S1} \rangle /(L_y\mathcal{L}_{eff})(S_e/S_n)$, where $\mathcal{L}_{eff}$ is the relative scintillation efficiency of nuclear recoils (also referred to as the effective Lindhard factor). A monoenergetic gamma (usually 122~keV), with a measured light yield $L_y$ given in photoelectrons/keV, is used as standard candle. An electric field applied across the xenon target allows the measurement of the electron signal, and it also quenches the scintillation signal. The quenching of the gamma is $S_e \sim 0.5$, while nuclear recoils are barely quenched ($S_n\sim0.95$). Both $L_y$ and $S_e$ vary significantly with the electric field strength.

\begin{figure}[h]
\begin{center}
\includegraphics[width=0.48\textwidth]{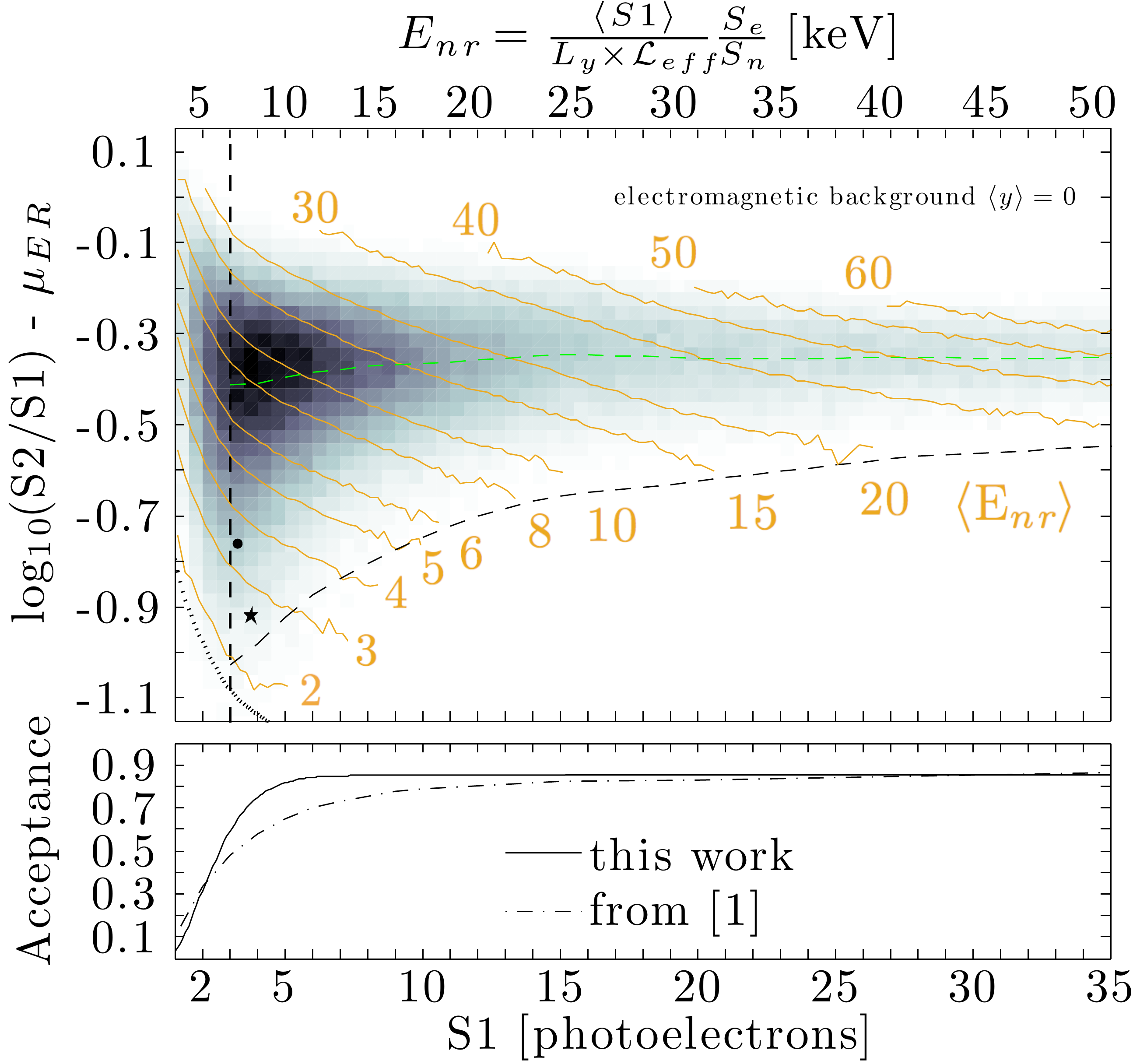}
\vskip -0.1cm
\caption{{\bf (upper panel)} Simulated nuclear recoil calibration data from a hypothetical detector which is similar to the XENON100 detector during its 2009 commissioning run. The dashed curves, described in the text, define three sides of the signal search box. The inset contours indicate $\langle {E}_{nr} \rangle$. Electron recoils from electromagnetic background have $\langle y \rangle=0$. The energy scale used in \cite{:2012nq} is indicated along the top, and the two events reported therein are reproduced for comparison. {\bf (lower panel)} S1 acceptance of the hypothetical detector (solid), and the XENON100 detector (dashed).}
\vskip -0.5cm
\label{fig:energyContours}
\end{center}
\end{figure}

Dedicated, direct measurements of the relative scintillation efficiency $\mathcal{L}_{eff}$ are plentiful \cite{Manzur:2009hp,Aprile:2008rc,Plante:2011hw}. These measurements report, for a given ${E}_{nr}$, the average number of recorded photoelectrons $\langle \mbox{S1} \rangle$. In contrast, dark matter search experiments report a measured S1 (and S2) corresponding to each event, and wish to know the most likely nuclear recoil energy $\langle {E}_{nr} \rangle$ associated with that event. For events near the nuclear recoil band centroid (in $y$), these two operations commute. But in many cases, such as the recently reported two events in the XENON100 search region \cite{:2012nq}, they do not.

It appears that a nuclear recoil energy reconstruction based directly on Lindhard theory is possible in liquid xenon \cite{Sorensen:2011bd}. The method employed in this work is very nearly equivalent to that, but framed so as to make an explicit connection with the historical method, and with the experimentally measured quantities. 

\begin{figure}[h]
\begin{center}
\includegraphics[width=0.48\textwidth]{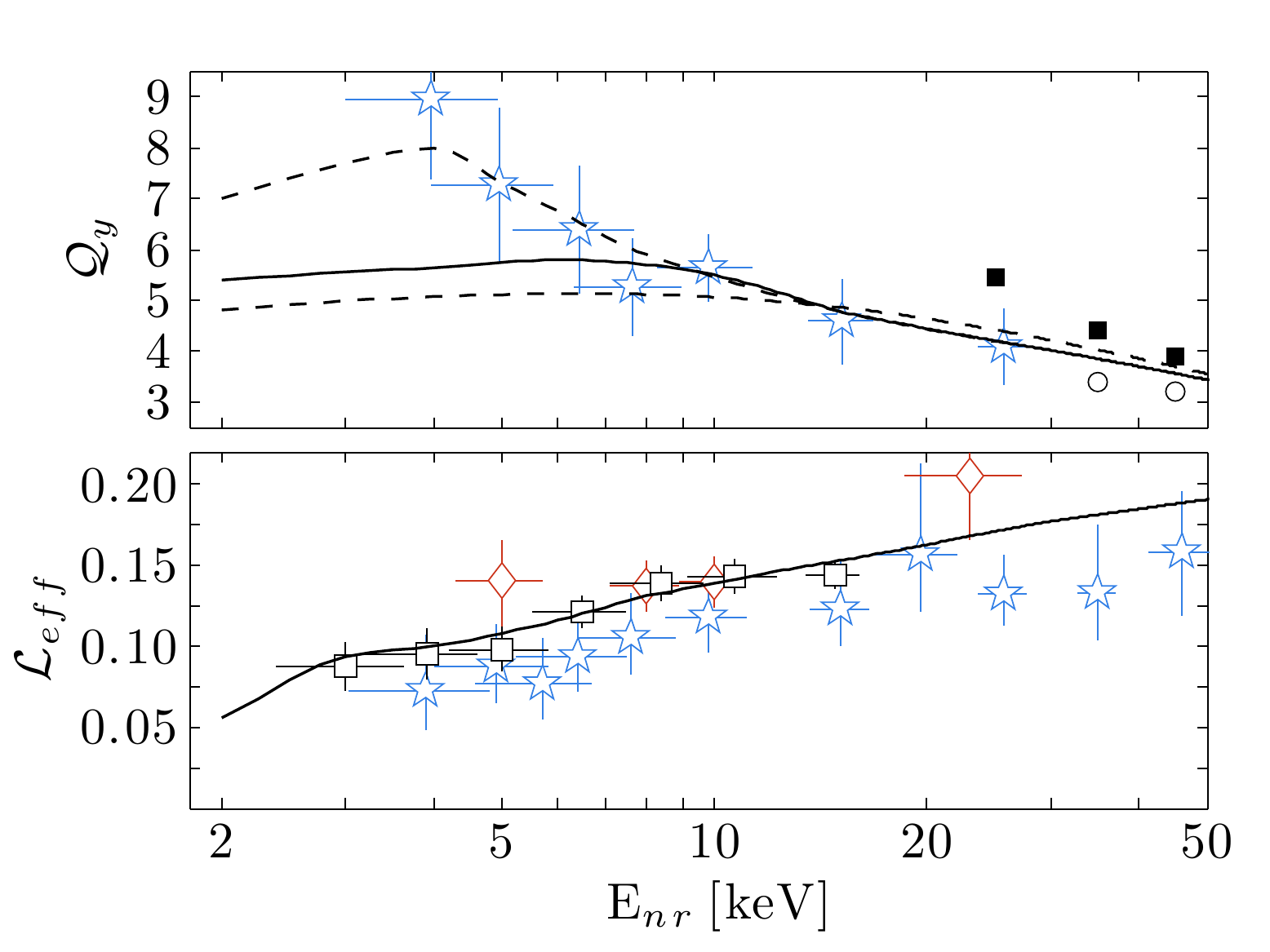}
\vskip -0.1cm
\caption{Nuclear recoil energy calibration data for liquid xenon. The simulation assumes the solid curves. Events with $E_{nr}<2$~keV were not simulated. $\mathcal{L}_{eff}$ data reproduced from \cite{Manzur:2009hp} (stars), \cite{Aprile:2008rc} (diamonds) and \cite{Plante:2011hw} (squares). Additional $\mathcal{Q}_y$ data from \cite{Aprile:2006kx}. } 
\vskip -0.5cm
\label{fig:energyScale}
\end{center}
\end{figure}

Figure \ref{fig:energyContours} shows the results of a simulation of the nuclear recoil response of a hypothetical liquid xenon detector. The results are plotted in the customary variables, though other choices are possible \cite{Arisaka:2012ce}, and may be useful considering the relative size of $\alpha_1$ and $\alpha_2$. The simulation method is described in \cite{Sorensen:2010hq},  and reproduces the relevant binomial and Poisson statistical processes. It has been shown to provide a hi-fidelity reproduction of XENON10 nuclear recoil calibration data. The detector-specific details were obtained from \cite{Aprile:2010um,Aprile:2011dd}. The hypothetical detector is therefore expected to exhibit a response similar to the XENON100 detector during its initial phase of operation. Uncertainties are discussed in a separate section. In this work, we took as inputs the nuclear recoil centroid reported in Fig. 3 of \cite{Aprile:2010um}, and the central $\mathcal{L}_{eff}$ curve from \cite{Aprile:2011hi} (shown in Fig. \ref{fig:energyScale}, lower panel, solid curve). The simulated data are shown after subtracting the electron recoil centroid ($\mu_{ER}$), which was also taken from \cite{Aprile:2010um}. 

The simulation has only a single free parameter, the ionization yield $\mathcal{Q}_y\equiv\mbox{S2}/{E}_{nr}$. It was allowed to float until the simulation band centroid matched the nuclear recoil centroid from data \cite{Aprile:2010um}. The agreement is very good, within $1\sigma$ of the statistical uncertainty on the mean, above $\mbox{S1}=3$. Below $\mbox{S1}=3$, the agreement is within $2\sigma$. The $\mathcal{Q}_y$ curve so obtained is shown in Fig. \ref{fig:energyScale} (solid curve). This does not guarantee that either $\mathcal{L}_{eff}$ or $\mathcal{Q}_y$ are correct in absolute terms, but rather, as drawn, are self-consistent with nuclear recoil band data. We note that the lower (dashed) $\mathcal{Q}_y$ curve is very consistent with the NEST model \cite{Szydagis:2011tk}. 

In addition to the simulated nuclear recoil data, Fig. \ref{fig:energyContours} shows the centroid $\mu_{NR}$ (dashed, green) and $\mu_{NR}-3\sigma_{NR}$ (dashed, black). Also shown are the ``software'' $\mbox{S1}>3$ (dashed) and $\mbox{S2}>150$ (stippled) thresholds, as in \cite{:2012nq}. The dashed curves define three walls of the dark matter search box for our hypothetical detector. In a similar (actual) search box, the XENON100 Collaboration recently reported the observation of two events, which are reproduced here. The stated event energies are 7.1 (circle) and 7.8 keV (star), and this appears very reasonable according to the  ${E}_{nr}\propto \mbox{S1}$ scale given along the top axis. 

\begin{figure}[ht]
\begin{center}
\includegraphics[width=0.48\textwidth]{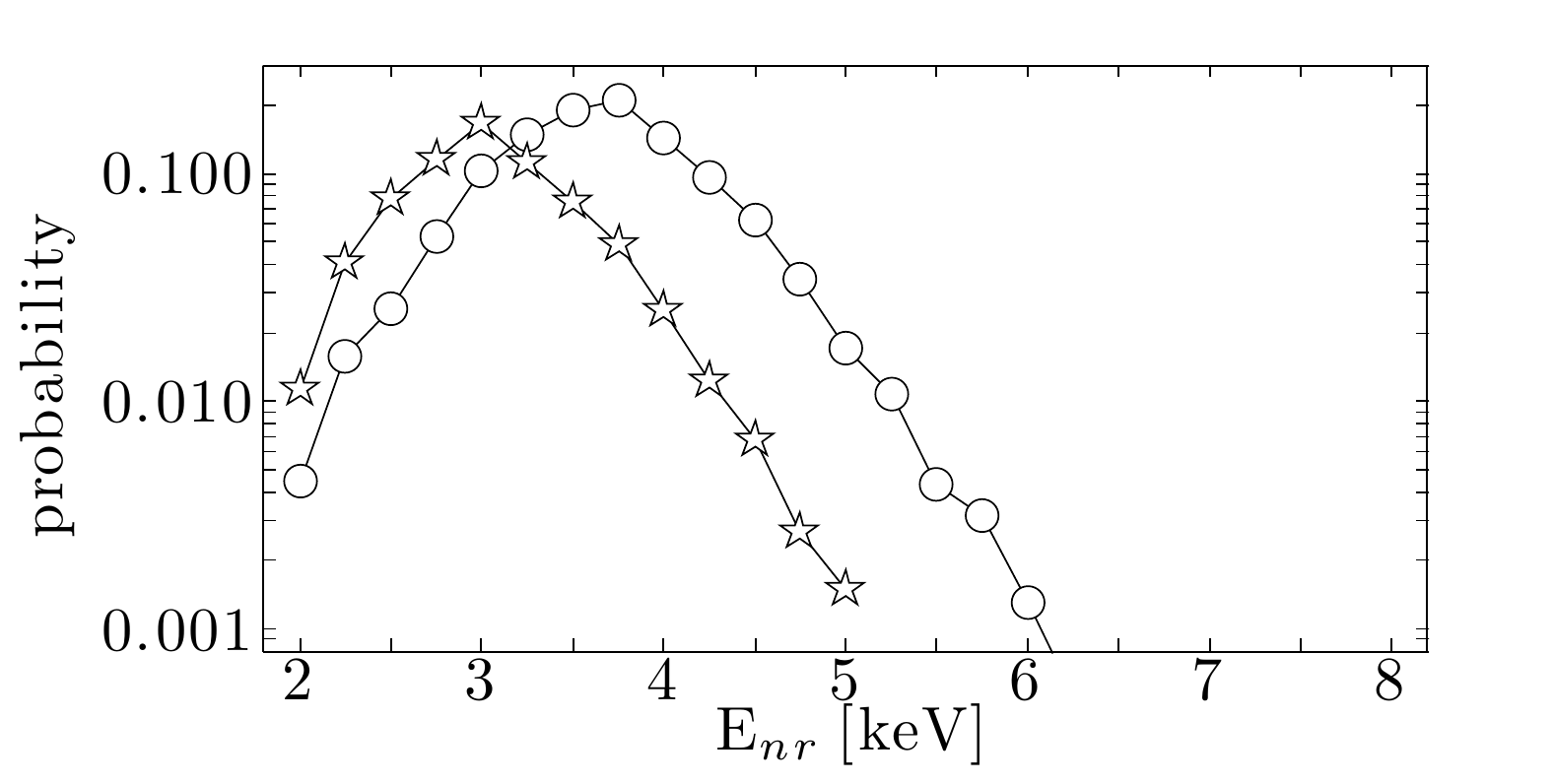}
\vskip -0.1cm
\caption{The probability to observe a fluctuation equal to or greater than the two events shown in Fig. \ref{fig:energyContours}, as a function of nuclear recoil energy. The abcissa is the actual simulated energy (not the contours in Fig. \ref{fig:energyContours}), and the markers correspond to the two events.}
\vskip -0.5cm
\label{fig:energyProbability}
\end{center}
\end{figure} 

We used the simulated events to find contours of  $\langle {E}_{nr} \rangle$, as shown in Fig. \ref{fig:energyContours} (solid curves, with corresponding energy in keV). Because a downward fluctuation in S1 is accompanied by an upward fluctuation in $y$, the contours follow the S2 expectation value for each energy. This is significantly different from the cartesian expectation implied by ${E}_{nr}\propto \mbox{S1}$ (notice that this scale is most correct near the calibration centroid, near $y\approx-0.4$). Figure \ref{fig:energyProbability} shows the probability that a nuclear recoil of energy $E_{nr}$ resulted in either of the two observed events. Only simulated events which produced a measurable S1 and S2 signal were considered.

\begin{figure*}[t]
\begin{center}
\includegraphics[width=0.98\textwidth]{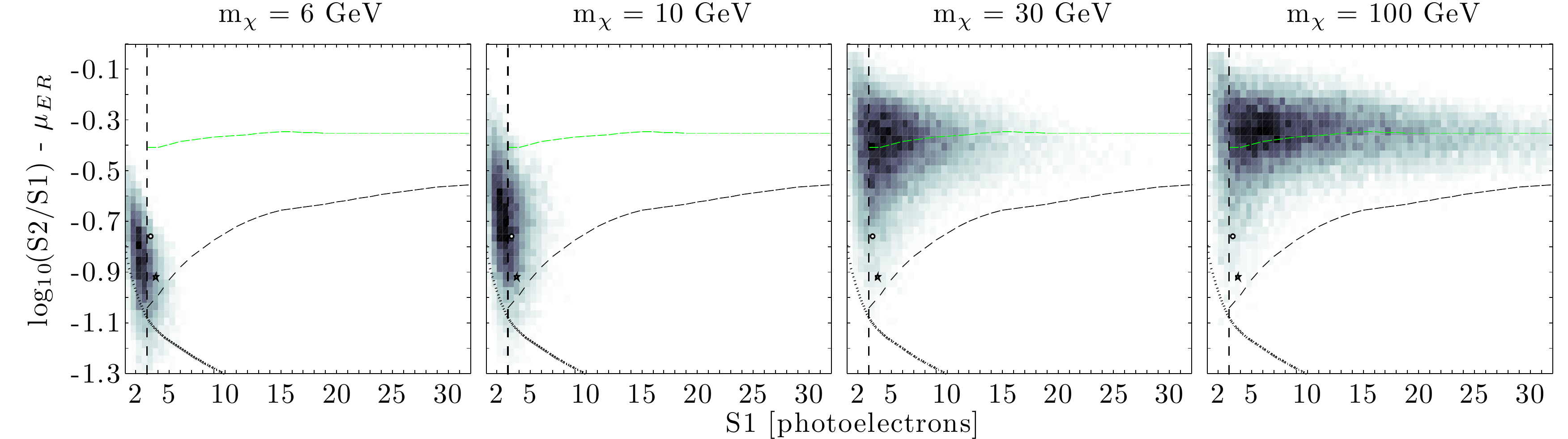}
\vskip -0.1cm
\caption{The expected distribution of nuclear recoil events, for several values of $m_{\chi}$. In order to clearly show the distribution, the following cross sections $\sigma_n$ were assumed: $5\times10^{-39}$, $1\times10^{-40}$, $1\times10^{-41}$ and $1\times10^{-41}$~cm$^2$.}
\vskip -0.5cm
\label{fig:examples}
\end{center}
\end{figure*} 

The discussion up to this point has assumed a nuclear recoil spectrum corresponding to an americium-berylium neutron source, as is frequently used to calibrate the nuclear recoil response of liquid xenon detectors. In Fig. \ref{fig:examples} we show the expected distributions of events for several dark matter masses $m_{\chi}$. The spectral distributions were calculated assuming the same astrophysical parameters described in \cite{:2012nq}. This clearly shows, particularly for $m_{\chi}\lesssim10$~GeV, that the $y$ coordinate also carries spectral information. The point is perhaps obvious from the definition of $y$, but it has been neglected in previous work. 

\mysection{{\it Uncertainties}}
\label{sec:uncert}
The ordinary statistical processes modeled by the simulation lead to  non-Gaussian tails in the distribution of $y$, particularly for $\mbox{S1}\lesssim10$. The extent of the tails are roughly indicated by the $\langle {E}_{nr} \rangle$ contours in Fig. \ref{fig:energyContours}. In Fig. \ref{fig:energyScale} (upper panel), we indicate the approximate range of $\mathcal{Q}_y$ (dashed curves) which adequately reproduce the nuclear recoil band data centroid in \cite{Aprile:2010um}, given $\mathcal{L}_{eff}$ as in the lower panel (solid curve). The lower dashed curve is the model prediction employed in \cite{Angle:2011th}; $\mathcal{Q}_y$ below this do not appear reasonable. It can be seen that the uncertainty in $\mathcal{Q}_y$ is most significant for ${E}_{nr}\lesssim10$~keV. This is due primarily to the non-Gaussian tails in $y$, and how the mean value of the distribution is determined (our method may be slightly different from what was used in \cite{Aprile:2010um}). As a result, one may expect a systematic shift in the $\langle {E}_{nr} \rangle$ contours in Fig. \ref{fig:energyContours}, of as much as $\Delta y = ^{+0.15}_{-0.06}$ for ${E}_{nr}\lesssim10$~keV. 

The S1 acceptance of our hypothetical detector response is different from  \cite{:2012nq}, as shown in Fig. \ref{fig:energyContours} (lower panel). The statistical methods we employ make it easy to adjust the predicted S1 acceptance to match that of \cite{:2012nq}. This has a systematic effect on the best-fit $\mathcal{Q}_y$, as pointed out in \cite{Aprile:2012he}, and leads to a maximum displacement $\Delta y = +0.02$ of the $\langle {E}_{nr} \rangle$ contours. This effect causes the largest shift in the range $3-5$~keV.

Of course, there is also uncertainty in $\mathcal{L}_{eff}$ itself. In this work, we have taken as a prior the $\mathcal{L}_{eff}$ favored by the XENON100 Collaboration. An equally plausible choice would be \cite{Manzur:2009hp} (shown in Fig. \ref{fig:energyScale}, stars). A 20\% smaller $\mathcal{L}_{eff}$ would require a systematic decrease in $\mathcal{Q}_y$ of about $8-12\%$. This would lead to an essentially uniform $-5\%$ systematic shift in the $\langle {E}_{nr} \rangle$ contours, as can be verified analytically. 

It is also notable that the S1 response reported in \cite{:2012nq} has improved by $\sim4\%$ and the S2 response has improved by $\sim14\%$, relative to \cite{Aprile:2010um} (and hence relative to our hypothetical detector). This leads to a uniform $+4\%$ systematic shift in the $\langle {E}_{nr} \rangle$ contours. We mention in passing that the width of the simulated band (in $y$) appears in good agreement with previously reported results \cite{Aprile:2010um,Aprile:2011hi}, but slightly narrower than \cite{:2012nq}.  In spite of these uncertainties, the fundamental profile of the $\langle {E}_{nr} \rangle$ contours remains as shown in Fig. \ref{fig:energyContours}.

\mysection{{\it Conclusions}}
\label{sec:conclus}
Full consideration of the energy information carried by both the S1 and S2 signals indicates that the energies of the two events observed by the XENON100 detector would be reconstructed at $2.9\pm0.5$~keV and $3.6\pm0.6$~keV in our hypothetical detector, subject to a systematic uncertainty of $^{+0.3}_{-1.0}$~keV. This is about a factor $\times2$ smaller than what one obtains from ${E}_{nr}~\propto~\mbox{S1}$. Interestingly, the approximate location of these events (near the lower left corner of a typical search box, rather than near the nuclear recoil centroid) is what one would expect for elastic scattering of low-mass ($m_{\chi}\lesssim10$~GeV) dark matter. This observation could be important from a phenomenological perspective ({\it e.g.} \cite{Hooper:2012ft,Hooper:2012cw}). 

It is evident from Fig. \ref{fig:examples} is that the acceptance of the search box (as a function of S1) depends on $m_{\chi}$. For the four example $m_{\chi}$ values, the fraction of events below the calibration centroid in the range $3\leq\mbox{S1}<10$ are 1.00, 0.99, 0.56 and 0.42. This is a result of the shape of the calculated dark matter energy spectrum.  The fraction is about 0.53 for the calibration data, slightly above 0.50 due to the non-Gaussian tails. Energy reconstruction based on $E_{nr} \propto \mbox{S1}$ leads to the assumption that the acceptance of the search box is always given by the calibration data. Relative to previously reported results, this should tend to strengthen the sensitivity to particle masses $m_{\chi}\lesssim35$~GeV, and weaken the sensitivity to larger masses. From an experimental point of view, it is interesting that for $m_{\chi}\lesssim10$~GeV, the electromagnetic background population with $\langle y \rangle = 0$ is essentially irrelevant. 

\begin{figure}[ht]
\begin{center}
\includegraphics[width=0.48\textwidth]{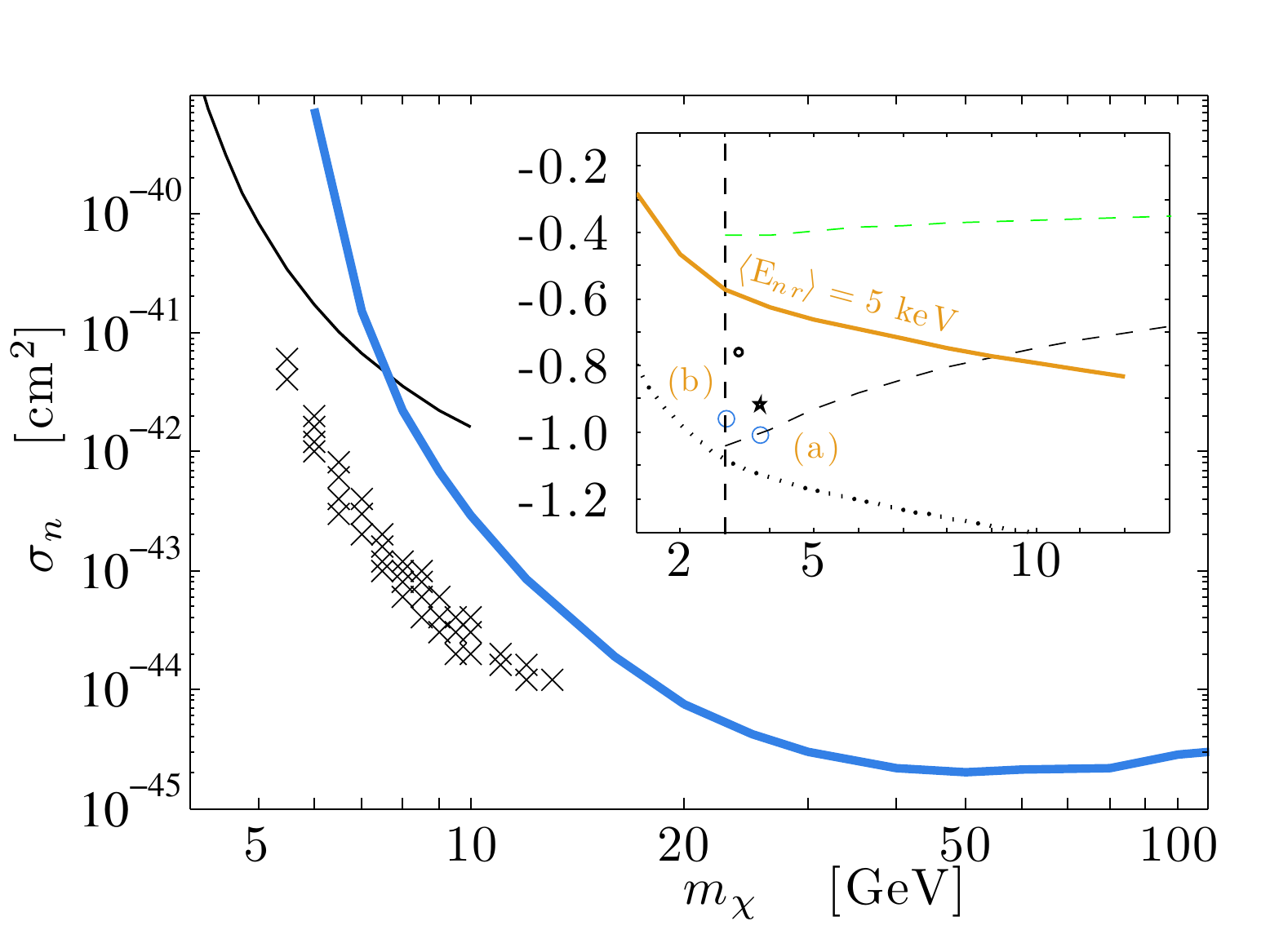}
\vskip -0.1cm
\caption{The region of parameter space in which an outcome similar to the two events shown in Fig. \ref{fig:energyContours} is at least 10\% likely (crosses), and the 90\% CL exclusion limits from XENON100 \cite{:2012nq} (thick curve) and XENON10 \cite{Angle:2011th} (thin curve).  {\bf (Inset)} the result of a simulated experiment, for $\sigma_n=3\times10^{-43}~\mbox{cm}^2$ and $m_{\chi}=7$~GeV. The two simulated events are shown as large circles; regions (a) and (b) are described in the text; markings and axes are as described in Fig. \ref{fig:energyContours}. }
\vskip -0.5cm
\label{fig:limits}
\end{center}
\end{figure} 

In Fig. \ref{fig:limits} we show values of $(m_{\chi} , \sigma_n)$, in which outcomes similar to that recently observed by  XENON100 appear at least 10\% likely (crosses) in our hypothetical detector. The S1 acceptance is shown in Fig. \ref{fig:energyContours} (solid curve), and a 34~kg~$\times$~224.6 day exposure was assumed. Our definition of ``similar'' is two events observed in the search box below the $\langle {E}_{nr} \rangle = 5$~keV contour, and no events above it. The search box is shown in Fig. \ref{fig:energyContours}, and Fig. \ref{fig:limits} (inset), bounded by dashed lines. Additionally, to account for the uncertainties discussed above, we allow (a) the possibility that one of the two events appears in the region below the search box, but above the S2 threshold, and (b) the possibility that one of the two events falls in the region S1$<3$. These regions are indicated in Fig. \ref{fig:limits} (inset). Uncertainties would propagate into Fig. \ref{fig:limits} as follows: a larger S1 acceptance ({\it e.g.} Fig. \ref{fig:energyContours} dashed curve) would tend to push the $(m_{\chi} , \sigma_n)$ region to smaller $\sigma_n$. A larger $\mathcal{Q}_y$ would mean a smaller energy for events near the lower left corner of the search box, and would therefore tend to push the $(m_{\chi} , \sigma_n)$ region to smaller $m_{\chi}$.

We do not suggest that these two events observed in \cite{:2012nq} are due to the elastic scattering of dark matter; the background hypothesis is of a similar likelihood, and thus more compelling. However, we have shown that in considering detection scenarios, significant additional information is gained from an energy reconstruction based on both $n_e$ and $n_{\gamma}$. Specifically, while a result similar to \cite{:2012nq} is compatible with low-mass dark matter, it is highly unlikely to have arisen from dark matter with $m_{\chi}\gtrsim10$~GeV. 


\begin{center} 
{\bf Acknowledgements}
\end{center}
Thanks are due to Adam Bernstein, Rouven Essig, Rick Gaitskell, Jeremy Mardon, Neal Weiner and the XENON100 Collaboration, for suggesting improvements to the manuscript. This work was performed under the auspices of the U.S. Department of Energy by Lawrence Livermore National Laboratory under contract DE-AC52-07NA27344. Report number LLNL-TR-574054.

{\it Note added.--} We thank the authors of \cite{Davis:2012hn} for bringing their related work to our attention.


\begin{thebibliography}{99}


\bibitem{:2012nq} 
  E.~Aprile {\it et al.}  [XENON100 Collaboration],
  arXiv:1207.5988 [astro-ph.CO].

\bibitem{Akerib:2011ix} 
  D.~S.~Akerib {\it et al.}  [LUX Collaboration],
  Nucl.\ Instrum.\ Meth.\ A {\bf 668}, 1 (2012)
  [arXiv:1108.1836 [astro-ph.IM]].
  
\bibitem{Gaitskell:2004gd} 
  R.~J.~Gaitskell,
  Ann.\ Rev.\ Nucl.\ Part.\ Sci.\  {\bf 54}, 315 (2004).
  
\bibitem{Chepel:2012sj} 
  V.~Chepel and H.~Araujo,
  arXiv:1207.2292 [physics.ins-det].

\bibitem{Sorensen:2011bd} 
  P.~Sorensen and C.~E.~Dahl,
  Phys.\ Rev.\ D {\bf 83}, 063501 (2011)
  [arXiv:1101.6080 [astro-ph.IM]].

\bibitem{Manzur:2009hp} 
  A.~Manzur, A.~Curioni, L.~Kastens, D.~N.~McKinsey, K.~Ni and T.~Wongjirad,
  Phys.\ Rev.\ C {\bf 81}, 025808 (2010)
  [arXiv:0909.1063 [physics.ins-det]].

\bibitem{Aprile:2008rc} 
  E.~Aprile, L.~Baudis, B.~Choi, K.~L.~Giboni, K.~Lim, A.~Manalaysay, M.~E.~Monzani and G.~Plante {\it et al.},
  Phys.\ Rev.\ C {\bf 79}, 045807 (2009)
  [arXiv:0810.0274 [astro-ph]].
  
\bibitem{Plante:2011hw} 
  G.~Plante, E.~Aprile, R.~Budnik, B.~Choi, K.~L.~Giboni, L.~W.~Goetzke, R.~F.~Lang and K.~E.~Lim {\it et al.},
  Phys.\ Rev.\ C {\bf 84}, 045805 (2011)
  [arXiv:1104.2587 [nucl-ex]].

\bibitem{Arisaka:2012ce} 
  K.~Arisaka, P.~Beltrame, C.~Ghag, K.~Lung and P.~R.~Scovell,
  arXiv:1202.1924 [astro-ph.IM].
  
\bibitem{Sorensen:2010hq} 
  P.~Sorensen,
  JCAP {\bf 1009}, 033 (2010)
  [arXiv:1007.3549 [astro-ph.IM]].
  
\bibitem{Aprile:2010um} 
  E.~Aprile {\it et al.}  [XENON100 Collaboration],
  Phys.\ Rev.\ Lett.\  {\bf 105}, 131302 (2010)
  [arXiv:1005.0380 [astro-ph.CO]].

\bibitem{Aprile:2011dd} 
  E.~Aprile {\it et al.}  [XENON100 Collaboration],
  Astropart.\ Phys.\  {\bf 35}, 573 (2012)
  [arXiv:1107.2155 [astro-ph.IM]].

\bibitem{Aprile:2011hi} 
  E.~Aprile {\it et al.}  [XENON100 Collaboration],
  Phys.\ Rev.\ Lett.\  {\bf 107}, 131302 (2011)
  [arXiv:1104.2549 [astro-ph.CO]].

\bibitem{Szydagis:2011tk} 
  M.~Szydagis, N.~Barry, K.~Kazkaz, J.~Mock, D.~Stolp, M.~Sweany, M.~Tripathi and S.~Uvarov {\it et al.},
  JINST {\bf 6}, P10002 (2011)
  [arXiv:1106.1613 [physics.ins-det]]; see also M.~Szydagis talk at \href{http://kicp-workshops.uchicago.edu/IDM2012/}{IDM 2012}, Chicago, USA.
  
\bibitem{Aprile:2006kx} 
  E.~Aprile, C.~E.~Dahl, L.~DeViveiros, R.~Gaitskell, K.~L.~Giboni, J.~Kwong, P.~Majewski and K.~Ni {\it et al.},
  Phys.\ Rev.\ Lett.\  {\bf 97}, 081302 (2006)
  [astro-ph/0601552].
  
\bibitem{Aprile:2012he} 
  E.~Aprile {\it et al.}  [The XENON Collaboration],
  arXiv:1208.5762 [astro-ph.CO].

\bibitem{Angle:2009xb} 
  J.~Angle {\it et al.}  [XENON10 Collaboration],
  Phys.\ Rev.\ D {\bf 80}, 115005 (2009)
  [arXiv:0910.3698 [astro-ph.CO]].

\bibitem{Aprile:2012vw} 
  E.~Aprile, M.~Alfonsi, K.~Arisaka, F.~Arneodo, C.~Balan, L.~Baudis, A.~Behrens and P.~Beltrame {\it et al.},
  arXiv:1207.3458 [astro-ph.IM].

\bibitem{Hooper:2012ft} 
  D.~Hooper,
  arXiv:1201.1303 [astro-ph.CO].
  
\bibitem{Hooper:2012cw} 
  D.~Hooper, N.~Weiner and W.~Xue,
  Phys.\ Rev.\ D {\bf 86}, 056009 (2012)
  arXiv:1206.2929 [hep-ph].

    
\bibitem{Angle:2011th} 
  J.~Angle {\it et al.}  [XENON10 Collaboration],
  Phys.\ Rev.\ Lett.\  {\bf 107}, 051301 (2011)
  [arXiv:1104.3088 [astro-ph.CO]].

\bibitem{Davis:2012hn} 
  J.~H.~Davis, T.~Ensslin and C.~Boehm,
  arXiv:1208.1850
  
\end{thebibliography}
\end{document}